\documentclass[12pt]{article}
\usepackage{amsmath}
\usepackage{amssymb}
\usepackage{epsfig}
\usepackage{euscript}
\usepackage{fancybox}
\usepackage{color}
\def\Color#1{\color[named]{#1}}

\def\mathswitchr#1{\relax\ifmmode{\mathrm{#1}}\else$\mathrm{#1}$\fi}

\newcommand{\umf}{{\Color{PineGreen}\mathfrak{u}}}
\newcommand {\pslash}{\hbox{$\not\hbox{\kern-2.3pt $p$}$}}

\usepackage{cite}

\usepackage{epic}

\def\alf1{ {\alpha\over\pi} }

\begin{document}
\begin{titlepage}
\begin{flushright}
{\bf BU-HEPP-07-08 }\\
{\bf June, 2007}\\
\end{flushright}
 
\begin{center}
{\Large IR-Improved DGLAP-CS Theory: Kernels, Parton Distributions, Reduced Cross Sections$^{\dagger}$
}
\end{center}

\vspace{2mm}
\begin{center}
{\bf   B.F.L. Ward}\\
\vspace{2mm}
{\em Department of Physics,\\
 Baylor University, Waco, Texas, 76798-7316, USA}\\
\end{center}

\vspace{5mm}
\begin{center}
{\bf   Abstract}
\end{center}
It is shown that exact, amplitude-based resummation allows IR-improvement
of the usual DGLAP-CS theory. This results in a new set of kernels, parton distributions and attendant reduced cross sections, so that the QCD perturbative
result for the respective hadron-hadron or lepton-hadron cross section
is unchanged order-by-order in $\alpha_s$ at large squared-momentum transfers. 
We compare these new objects with their usual counter-parts and illustrate
the effects of the IR-improvement in some phenomenological cases of interest
with an eye toward precision applications in LHC physics scenarios.
 
\par 
\vspace{10mm}
\vspace{10mm}
\renewcommand{\baselinestretch}{0.1}
\footnoterule
\noindent
{\footnotesize
\begin{itemize}
\item[${\dagger}$]
Work partly supported by US DOE grant DE-FG02-05ER41399 and
by NATO grant PST.CLG.980342.
\end{itemize}
}

\end{titlepage}

\def\Kmax{K_{\rm max}}\def\ieps{{i\epsilon}}\def\rQCD{{\rm QCD}}
\renewcommand{\theequation}{\arabic{equation}}
\font\fortssbx=cmssbx10 scaled \magstep2
\renewcommand\thepage{}
\parskip.1truein\parindent=20pt\pagenumbering{arabic}\par
\section{\bf Introduction}\label{intro}\par
With the impending start of the LHC, the era of precision QCD at the LHC, by which we mean 1\% or better precision tags on the theoretical predictions, 
obtains. In this new era for perturbative QCD in high energy colliding beam physics, we have the 
extremely challenging task of proving that a given theoretical precision tag
does in fact hold to that level. All aspects 
of the standard formula
for hadron-hadron scattering in perturbative QCD have to be examined
for possible sources of uncertainty in the physical and technical precision
components of any claimed total theoretical precision tag.
In this connection, we recall that, in Ref.~\cite{irdglap}, we have 
found that the resummation of large infrared (IR) effects in the kernels of
the usual DGLAP-CS~\cite{c-s,pgp,djgfw,dglap} theory results in their improved 
behavior in the respective IR limit. This improved behavior,
for example, results in kernels that are integrable in the IR limit
and which therefore are more amenable to realization by MC methods 
(our ultimate goal) to
arbitrary precision. In this paper, we review the 
results in Ref.~\cite{irdglap}
and we show that the new IR-improved DGLAP-CS theory leads to a new set of
parton distributions and reduced cross sections,
so that the exact predictions from QCD for hadron-hadron and lepton-hadron
processes are unaltered order-by-order in
perturbation theory. The advantage is better control on the accuracy
of a given fixed-order calculation throughout the entire phase space
of the respective physical process, especially when the prediction
is given by MC methods.
\par
The paper is organized as follows. In the next section, we recapitulate
the new IR-improved DGLAP-CS theory, as it is not very familiar. In Section 3,
we illustrate the use of the new IR-improved theory with some
examples of phenomenological interest. In Section 4,
we address the issue of what it does to the standard calculational apparatus
for perturbative QCD predictions of hadron-hadron 
and lepton-hadron scattering processes
at large squared momentum transfer.
Section 5 contains some concluding
remarks. Review of exact, amplitude-based resummation theory
is presented in the Appendix.
\par
\section{IR-Improved DGLAP-CS Theory}
 
Specifically, the motivation for the improvement which we develop
can be seen already in the basic results in Refs.~\cite{dglap}
for the kernels that determine the evolution of the structure
functions by the attendant DGLAP-CS evolution of the corresponding
parton densities by the standard methodology. Here, we already stress
that the attendant evolution equations, 
under Mellin tranformation, are entirely
implied by those of the Callan-Symanzik-type~\cite{c-s} 
analyzed in Refs.~\cite{pgp,djgfw}
in their classic analysis of the deep inelastic scattering processes.
Thus, henceforward, we shall refer to these equations as 
the DGLAP-CS equations.
Consider the evolution of the
non-singlet(NS) parton density function $q^{NS}(x)$, where $x$
can be identified as Bjorken's variable as usual.
The basic starting point of our analysis is the infrared divergence
in the kernel that determines this evolution:
\begin{equation}
\frac{dq^{NS}(x,t)}{dt}=\frac{\alpha_s(t)}{2\pi}\int_{x}^{1}\frac{dy}{y}q^{NS}(y,t)P_{qq}(x/y)
\label{dglap1}
\end{equation}
where the well-known result for the kernel $P_{qq}(z)$ is,
for $z<1$,
\begin{equation}
P_{qq}(z)= C_F\frac{1+z^2}{1-z}
\label{dglap2}
\end{equation}
when we set $t=\ln \mu^2/\mu_0^2$ for some reference scale $\mu_0$
with which we study evolution to the scale of interest $\mu$.
\footnote{We will generally follow Ref.~\cite{field} and set
$\mu_0=\Lambda_{QCD}$ without loss of content since $dt=dt'$
when $t=\ln\mu^2/\Lambda_{QCD}^2,~ t'=\ln\mu^2/\mu_0^2$ for fixed
values of $\Lambda_{QCD},\mu_0$.}
Here, $C_F=(N_c^2-1)/(2N_c)$ is the quark color 
representation's quadratic Casimir invariant
where $N_c$ is the number of colors and so that it is just 3. 
This kernel has an unintegrable IR singularity at $z=1$, which is the point of
zero energy gluon emission and this is as it should be.
The standard treatment of this very physical effect is to
regularize it by the replacement
\begin{equation}
\frac{1}{(1-z)}\rightarrow \frac{1}{(1-z)_+}
\label{dglap3}
\end{equation} 
with the distribution $\frac{1}{(1-z)_+}$ defined so that
for any suitable test function $f(z)$ we have
\begin{equation}
\int_{0}^{1}dz\frac{f(z)}{(1-z)_+}=\int_{0}^{1}dz\frac{f(z)-f(1)}{(1-z)}.
\label{dglap4}
\end{equation}
A possible representation of $1/(1-z)_+$ is seen to be
\begin{equation}
\frac{1}{(1-z)_+}=\frac{1}{(1-z)}\theta(1-\epsilon-z)+\ln\epsilon\,\delta(1-z)
\label{dglap5}
\end{equation}
with the understanding that $\epsilon\downarrow 0$.
We use the notation $\theta(x)$ for the step function from $0$ for $x<0$
to $1$ for $x\ge 0$ and $\delta(x)$ is Dirac's delta function.
The final result for $P_{qq}(z)$ is then obtained by imposing the
physical requirement~\cite{dglap} that
\begin{equation}
\int_{0}^{1}dzP_{qq}(z)=0,
\label{dglap6}
\end{equation}  
which is satisfied by adding the effects of virtual corrections at
$z=1$ so that finally
\begin{equation}
P_{qq}(z)= C_F\left( \frac{1+z^2}{(1-z)_+}+\frac{3}{2}\delta(1-z)\right). 
\label{dglap7}
\end{equation}  
\par

The smooth behavior in the original real emission 
result from the Feynman rules,
with a divergent $1/(1-z)$ behavior as $z\rightarrow 1$, has been replaced
with a mathematical artifact: the regime $1-\epsilon< z<1$ now has
no probability at all 
and at $z=1$ we have a large negative integrable contribution
so that we end-up finally with a finite (zero) value for the 
total integral of $P_{qq}(z)$. This mathematical artifact is what we wish
to improve here; for, in the precision studies of 
Z physics~\cite{yellowbook,jsw,jw} at LEP1,
it has been found that such mathematical artifacts can indeed impair
the precision tag which one can achieve with a given fixed order
of perturbation theory. An analogous case is now well-known in the theory of
QCD higher order
corrections, where the FNAL data on $p_T$ spectra clearly show the 
need for improvement of fixed-order results by resumming large logs associated with soft gluons~\cite{cosopster,berge,cdf1}.\footnote{Note that we do not address $p_T$ resummation effects herein; we use this example to illustrate the point that, whenever the phase space for radiation is restricted in a given variable, soft gluon radiation will result in large logs that need to be resummed, 
altough the methods to make this resummation may vary in detail.} 
For reference, note that at the LHC, 2 TeV
partons are realistic so that $z\cong 0.001$ means $\sim 2-3$ GeV soft gluons, which are clearly above the LHC detector thresholds, 
in complete analogy with the situation
at LEP where $z\cong 0.001$ meant $\sim 100$ MeV photons which 
were also above the LEP detector thresholds -- just as resummation 
was necessary to describe this view of the LEP data, so too we may 
argue it will be necessary to describe the LHC data on the corresponding view.
And, more importantly, why should we have
to set $P_{qq}(z)$ to $0$ for $1-\epsilon< z < 1$ when it actually has
its largest values in this very regime?
\par

By mathematical artifact we do not mean that there is an error 
in the computations that lead to it; indeed, it is well-known
that this +-function behavior is exactly what one gets at
${\cal O}(\alpha_s)$ for the bremsstrahlung process. The artifact
is that the behavior of the differential spectrum of 
the process for $z\rightarrow 1$ in ${\cal O}(\alpha_s)$
is unintegrable and has to be cut-off 
and thus this spectrum is only poorly represented by the ${\cal O}(\alpha_s)$ calculation; for, the resummation of the large soft higher order effects as 
we present below changes the $z\rightarrow 1$ behavior non-trivially,
as from our resummation we will find that the $\frac{1}{1-z}$-behavior
is modified to $(1-z)^{\gamma-1}, \gamma>0$.
This is a testable effect, as we have seen in its QED analogs in Z physics at
LEP1~\cite{yellowbook,jsw,jw}: 
if the experimentalist measures the cross section
for bremsstrahlung for gluons(photons) down to energy fraction $\epsilon_0,~\epsilon_0>0$, in our new resummed theory presented below, 
the result will approach
a finite value from below as $-\epsilon_0^\gamma$ whereas the ${\cal O}(\alpha_s)$ +-function prediction would increase without limit as $-\ln\epsilon_0$.
The exponentiated result has been verified by the data at LEP1.\par

The important point is that the traditional resummations in N-moment
space for the DGLAP-CS kernels address only the short-distance 
contributions to their higher order corrections. The deep question we
deal with in this paper concerns, then, how much of the complete
soft limit of the DGLAP-CS kernels is contained in the anomalous dimensions
of the leading twist operators in Wilson's expansion, an expansion
which resides on the very tip of the light-cone? Are all of the effects of
the very soft gluon emission, involving, as they most certainly do, arbitrarily
long wavelength quanta, representable by the physics
at the tip of the light-cone? The Heisenberg uncertainty principle
surely tells us that answer can not be affirmative. In this paper, we
calculate these long-wavelength gluon effects on the DGLAP-CS kernels
that are not included (see the discussion below)
in the standard treatment of Wilson's expansion. We therefore do not
contradict the results of the large N-moment space resummations 
such as that presented in Ref.~\cite{alball} nor do we
contradict the renormalon chain-type resummation as done in
Ref.~\cite{mikhlv}.\par

We employ the exact re-arrangement of the Feynman series for QCD 
as it has been shown in Ref.~\cite{qcdexp,delaney}. For completeness,
as this QCD exponentiation theory is not generally familiar, we
reproduce its essential aspects in our Appendix. The idea is to sum up the
leading IR terms in the corrections to $P_{qq}$ with the goal that
they will render integrable the IR singularity that we have in its lowest order
form. This will remove the need for mathematical artifacts
and exhibit more accurately the true predictions of the 
full QCD theory in terms of fully
physical results.\par

\begin{figure}
\begin{center}
\setlength{\unitlength}{1mm}
\begin{picture}(160,80)
\put(-2.4, -10){\makebox(0,0)[lb]{
\epsfig{file=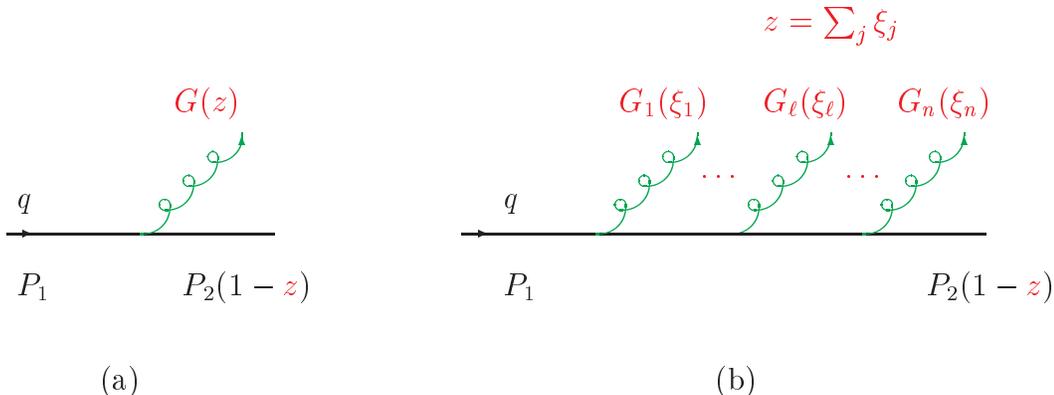,width=140mm}
}}
\end{picture}
\end{center}
\label{fig1-a}
\caption{In (a), we show the usual process $q\rightarrow q(1-z)+G(z)$;
in (b), we show its multiple gluon improvement $q\rightarrow q(1-z)+G_1(\xi_1)+\cdots+G_n(\xi_n),~~z=\sum_j\xi_j$.} 
\end{figure}

We apply the QCD exponentiation master formula in eq.(\ref{subp15})
in our Appendix
(see also Ref.~\cite{qcdexp}), following the
analogous discussion then for QED in Refs.~\cite{jsw,jw},
to the gluon emission transition that
corresponds to $P_{qq}(z)$, i.e., to the squared amplitude for
$q\rightarrow q(z)+G(1-z)$ so that in the Appendix one replaces
everywhere the squared amplitudes for the $\bar{Q}'Q\rightarrow \bar{Q}'''Q''$
processes with those for the former one plus its $nG$ analoga 
with the attendant changes in the
phase space and kinematics dictated by the standard methods; this implies
that in eq.(53) of the first paper in Ref.~\cite{dglap} we
have from eq.(\ref{subp15}) the replacement ( see Fig.~\ref{fig1-a} )
\noindent
\begin{equation}
\begin{split}
P_{BA}&=P_{BA}^0\equiv\frac{1}{2}z(1-z)\overline{\underset{spins}{\sum}}~\frac{|V_{A\rightarrow B+C}|^2}{p_\perp^2}\\
&\Rightarrow\\
P_{BA}&=\frac{1}{2}z(1-z)\overline{\underset{spins}{\sum}}~\frac{|V_{A\rightarrow B+C}|^2}{p_\perp^2}z^{\gamma_q}F_{YFS}(\gamma_q)e^{\frac{1}{2}\delta_q} 
\end{split}
\label{expn1-dglp}
\end{equation}
where $A=q$, $B=G$, $C=q$ and $V_{A\rightarrow B+C}$ is the lowest order
amplitude
for $q\rightarrow G(z)+q(1-z)$, so that 
we get the un-normalized exponentiated result
\begin{equation}
P_{qq}(z)= C_F F_{YFS}(\gamma_q)e^{\frac{1}{2}\delta_q}\frac{1+z^2}{1-z}(1-z)^{\gamma_q}
\label{dglap8}
\end{equation} 
where~\cite{qcdexp,delaney,jsw,jw} 
\begin{align}
\gamma_q &= C_F\frac{\alpha_s}{\pi}t=\frac{4C_F}{\beta_0}\\
\delta_q&=\frac{\gamma_q}{2}+\frac{\alpha_sC_F}{\pi}(\frac{\pi^2}{3}-\frac{1}{2})
\label{dglap9}
\end{align}
and 
\begin{equation}
F_{YFS}(\gamma_q)=\frac{e^{-C_E\gamma_q}}{\Gamma(1+\gamma_q)}.
\label{dglap10}
\end{equation}
Here, \[\beta_0=11-\frac{2}{3}n_f\], where $n_f$ is the number of
active quark flavors, \[C_E=.5772\dots\] is Euler's constant
and $\Gamma(w)$ is Euler's gamma function. The function
$F_{YFS}(z)$ was already introduced by Yennie, Frautschi
and Suura~\cite{yfs} in their analysis of the IR behavior
of QED. We see immediately that
the exponentiation has removed the unintegrable IR divergence at $z=1$.
For reference, we note that we have in (\ref{dglap8}) resummed
the terms\footnote{Following the standard LEP Yellow Book~\cite{yellowbook} convention, we do not include the order of the first nonzero term in
counting the order of its higher order corrections.} ${\cal O}(\ln^k(1-z)t^{\ell}\alpha_s^n),~~n\ge\ell\ge k$, which originate
in the IR regime and which exponentiate. The important point is that
we have not dropped outright the terms that do not exponentiate
but have organized them into the residuals $\tilde{\bar\beta}_m$
in the analog of eq.(\ref{subp15}).
   The application of eq.(\ref{subp15}) to obtain eq.(\ref{dglap8})
proceeds as follows. First, the exponent in the exponential factor in front of the expression on the RHS of eq.(\ref{subp15}) is readily seen to be from 
eq.(\ref{subp12}), using the well-known results for the respective
real and virtual infrared functions from Refs.~\cite{qcdexp,delaney},
\begin{equation}
\begin{split}
{SUM}_{IR}(QCD)&=2\alpha_s ReB_{QCD}+2\alpha_s\tilde B_{QCD}(\Kmax)\\
&=\frac{1}{2}\left(2 C_F\frac{\alpha_s}{\pi}t\ln{\frac{K_{max}}{E}}+ C_F\frac{\alpha_s}{2\pi}t+\frac{\alpha_sC_F}{\pi}(\frac{\pi^2}{3}-\frac{1}{2})\right)
\end{split}
\end{equation} 
where on the RHS of the last result we have already applied the DGLAP-CS
synthesization procedure in the third paper in Ref.~\cite{qcdexp} 
to remove the collinear
singularities, $\ln \Lambda^2_{QCD}/m_q^2 -1 $, in accordance with the standard
QCD factorization theorems~\cite{qcdfactorzn}. This means that, identifying
the LHS of eq.(\ref{subp15}) as the sum over final states and average 
over initial states of the respective process divided by the incident
flux and replacing that incident flux by the respective initial state
density according to the standard methods for the
process $q\rightarrow q(1-z)+G(z)$, occurring in the context
of a hard scattering at scale $Q$ as it is for eq.(53) in the first paper
in Ref.~\cite{dglap}, the soft gluon effects for energy fraction
$<z\equiv K_{max}/E$
give the result, from eq.(\ref{subp15}), that, working through to
the $\tilde{\bar\beta_1}$-level and using $q_2$ to represent the momentum
conservation via the other degrees of freedom for the attendant hard process, 
\begin{equation}
\begin{split}
\int\frac{\alpha_s(t)}{2\pi}P_{BA}dtdz&=e^{\rm SUM_{IR}(QCD)(z)}\int\{\tilde{\bar\beta}_0\int{d^4y\over(2\pi)^4}e^{\{iy\cdot(p_1-p_2)+\int^{k<K_{max}}{d^3k\over k}\tilde S_\rQCD(k)
\left[e^{-iy\cdot k}-1\right]\}}\\
&+ \int{d^3
k_1\over k_1}\tilde{\bar\beta}_1(k_1)\int{d^4y\over(2\pi)^4}e^{\{iy\cdot(p_1-p_2-k_1)+\int^{k<K_{max}}{d^3k\over k}\tilde S_\rQCD(k)
\left[e^{-iy\cdot k}-1\right]\}}\\
&+\cdots \}{d^3p_2\over p_2^{\,0}}{d^3q_2\over q_2^{\,0}} \\
&= e^{\rm SUM_{IR}(QCD)(z)}\int\{\tilde{\bar\beta}_0\int_{-\infty}^{\infty}{dy\over(2\pi)}e^{\{iy\cdot(E_1-E_2)+\int^{k<K_{max}}{d^3k\over k}\tilde S_\rQCD(k)
\left[e^{-iyk}-1\right]\}}\\
&+ \int{d^3
k_1\over k_1}\tilde{\bar\beta}_1(k_1)\int_{-\infty}^{\infty}{dy\over(2\pi)}e^{\{iy\cdot(E_1-E_2-k_1^0)+\int^{k<K_{max}}{d^3k\over k}\tilde S_\rQCD(k)
\left[e^{-iy\cdot k}-1\right]\}}\\
&+\cdots \}{d^3p_2\over p_2^{\,0}q_2^{\,0}} 
\end{split}
\label{expker1}
\end{equation}
where we set $E_i=p_i^0,~ i=1,2$ and the real infrared function $\tilde S_\rQCD(k)$ is well-known as well:
\begin{equation}
\tilde S_\rQCD(k)= -\frac{\alpha_s C_F}{8\pi^2}\left(\frac{p_1}{kp_1}-\frac{p_2}{kp_2}\right)^2|_{\text{DGLAP-CS synthesized}}
\label{realir}
\end{equation}
and we indicate as above that the DGLAP-CS synthesization
procedure in Refs.~\cite{qcdexp} is to be applied to its evaluation to
remove its collinear singularities; we are using the 
kinematics of the first paper in Ref.~\cite{dglap} in their
computation of $P_{BA}(z)$ in their eq.(53), so that the 
relevant value of $k_\perp^2$ is indeed $Q^2$. It means that  
the computation can also be seen to correspond to computing the IR 
function for the standard t-channel kinematics and taking $\frac{1}{2}$ 
of the result to match the single line emission in $P_{Gq}$. 
The two important integrals needed in (\ref{expker1}) were already
studied in Ref.~\cite{yfs}:
\begin{equation}
\begin{split}
I_{YFS}(zE,0)&=\int_{-\infty}^{\infty}\frac{dy}{2\pi}e^{[iy(zE)+\int^{k<zE}\frac{d^3k}{k}\tilde{S}_{QCD}(k)(e^{-iyk}-1)]}\\
&= F_{YFS}(\gamma_q)\frac{\gamma_q}{zE}\\
I_{YFS}(zE,k_1)&=\int_{-\infty}^{\infty}\frac{dy}{2\pi}e^{[iy(zE-k_1)+\int^{k<zE}\frac{d^3k}{k}\tilde{S}_{QCD}(k)(e^{-iyk}-1)]}\\
&=(\frac{zE}{zE-k_1})^{1-\gamma_q}I_{YFS}(zE,0)
\end{split}
\label{yfsintgls} 
\end{equation}
\par
When we introduce the results in (\ref{yfsintgls}) into (\ref{expker1})
we can identify the factor
\begin{equation}
\int\left(\tilde{\bar\beta}_0\frac{\gamma_q}{zE}+\int dk_1k_1d\Omega_1\tilde{\bar\beta}_1(k_1)(\frac{zE}{zE-k_1})^{1-\gamma_q}\frac{\gamma_q}{zE}\right)\frac{d^3p_2}{E_2q_2^{\,0}}=\int dt \frac{\alpha_s(t)}{2\pi}P_{BA}^0dz+{\cal O}(\alpha_s^2).
\label{1storder}
\end{equation}
where $P_{BA}^0$ is the unexponentiated result in the first line of
(\ref{expn1-dglp}). This leads us finally to the exponentiated 
result in the second line
of (\ref{expn1-dglp}) by elementary differentiation:
\begin{equation}
P_{BA}=P_{BA}^0z^{\gamma_q}F_{YFS}(\gamma_q)e^{\frac{1}{2}\delta_q}
\end{equation}
\par
Let us stress the following. In this paper, we have retained for pedagogical
reasons the dominant terms in the resummation which we use for the kernels.
The result in the first line of (\ref{expker1}) is exact and can be used
to include all higher order resummation effects systematically as desired.
Moreover, we have taken a one-loop representation of $\alpha_s$ for illustration and have set it to a fixed-value on the RHS of (\ref{expker1}), 
so that, thereby, we are dropping further possible subleading higher order effects, again for reasons of pedagogy. It is straight forward to include these 
effects as well~\cite{elswh}.
\par
Here, we also may note
how one can see that the terms we exponentiate are not included
in the standard treatment of Wilson's expansion: From the 
standard method~\cite{fyodor}, the N-th moment of the invariants $T_{i,\ell},~i=L,2,3,~\ell=q,G,$
of the forward Compton amplitude in DIS is projected by
\begin{equation}
{\cal P}_N\equiv \left[\frac{q^{\{\mu_1}\cdots q^{\mu_N\}}}{N!}\frac{\partial^N}{\partial{p^{\mu_1}}\cdots \partial{p^{\mu_N}}}\right]|_{p=0}
\end{equation}
where $x_{Bj}=Q^2/(2qp)$ in the standard DIS notation; this projects
the coefficient of $1/(2x_{Bj})^N$. For the dominant terms which we
resum here, the characteristic behavior would correspond formally
to $\gamma_q$-dependent anomalous dimensions associated with
the respective coefficient whereas by definition Wilson's
expansion does not contain such. In more phenomenologically familiar
language, it is well-known that the parton model used in this paper
to calculate the large distance effects that improve the kernels contains
such effects whereas Wilson's expansion does not: for example, the 
parton model can be used for Drell-Yan processes, Wilson's expansion
can not. Similarly, any Wilson-expansion guided procedure
used to infer the kernels via inverse Mellin transformation,
by calculating the coefficient of $(1/z)^n$ in Wilson's expansion,
will necessarily omit the dominant IR terms which we resum.
Here, we stress that we refer to the properties of the 
expansion of the invariant functions $T_i$, 
{\em not to the expansion of the kernels themselves}, 
as the latter are related to the respective 
anomalous dimension matrix elements by inverse Mellin transformations.
We need to emphasize that we are not saying that there is an error in
the standard use of Wilson's expansion. We are saying that the IR improvement
which we exhibit is not a property of that expansion as it is conventionally
defined.
\par
 
The normalization condition in eq.(\ref{dglap6}) then gives us the final
expression
\begin{equation}
P_{qq}(z)= C_F F_{YFS}(\gamma_q)e^{\frac{1}{2}\delta_q}\left[\frac{1+z^2}{1-z}(1-z)^{\gamma_q} -f_q(\gamma_q)\delta(1-z)\right]
\label{dglap11}
\end{equation} 
where
\begin{equation}
f_q(\gamma_q)=\frac{2}{\gamma_q}-\frac{2}{\gamma_q+1}+\frac{1}{\gamma_q+2}.
\label{dglap12}
\end{equation} 
The latter result is then our IR-improved kernel for NS DGLAP-CS
evolution in QCD. We note that the appearance of the integrable
function $(1-z)^{-1+\gamma_q}$ in the place of $\frac{1}{(1-z)_+}$
was already anticipated by Gribov and Lipatov in Refs.~\cite{dglap}.
Here, we have calculated the value of $\gamma_q$ in a systematic
rearrangement of the QCD perturbation theory that allows 
one to work to any exact order in the theory without
dropping any part of the theory's perturbation series.\par

The standard DGLAP-CS theory tells us that the kernel $P_{Gq}(z)$ is 
related to $P_{qq}(1-z)$ directly: for $z<1$, we have
\begin{equation}
P_{Gq}(z)=P_{qq}(1-z)= C_F F_{YFS}(\gamma_q)e^{\frac{1}{2}\delta_q}\frac{1+(1-z)^2}{z} z^{\gamma_q} .
\label{dglap13}
\end{equation}
This then brings us to our first non-trivial check of the new IR-improved theory; for, the conservation of momentum tells us that
\begin{equation}
\int_{0}^{1}dz z \left(P_{Gq}(z)+P_{qq}(z)\right) = 0.
\label{dglap14}
\end{equation}
Using the new results in eqs.(\ref{dglap11},\ref{dglap13}), we have
to check that the following integral vanishes:
\begin{equation}
I = \int_{0}^{1}dz z \left(\frac{1+(1-z)^2}{z} z^{\gamma_q}+\frac{1+z^2}{1-z}(1-z)^{\gamma_q} -f_q(\gamma_q) \delta(1-z)\right).
\label{dglap15}
\end{equation}
To see that it does, note that
\begin{equation}
\frac{z}{1-z}=\frac{z-1+1}{1-z}=-1+\frac{1}{1-z}.
\end{equation}
Introducing this result into eq.(\ref{dglap15}) we get
\begin{equation}
I = \int_{0}^{1}dz\{ (1+(1-z)^2)z^{\gamma_q}-(1+z^2)(1-z)^{\gamma_q}+\frac{1+z^2}{1-z}(1-z)^{\gamma_q} -f_q(\gamma_q) \delta(1-z)\}.
\label{dglap16}
\end{equation}
The integrals over the first two terms on the right-hand side (RHS)
of (\ref{dglap16}) exactly cancel as one sees by using the change
of variable $z\rightarrow 1-z$ in one of them and the integral
over the last two terms on the RHS of (\ref{dglap16}) vanishes from the
normalization in eq.(\ref{dglap6}). Thus we conclude that
\begin{equation}
I=0.
\end{equation}
The quark momentum sum rule is indeed satisfied.
\par

Having improved the IR divergence properties of $P_{qq}(z)$ and
$P_{Gq}(z)$, we now turn to $P_{GG}(z)$ and $P_{qG}(z)$.
We first note that the standard formula for $P_{qG}(z)$,
\begin{equation}
P_{qG}(z)=\frac{1}{2}(z^2+(1-z)^2),
\label{dglap17a}
\end{equation}
is already well-behaved (integrable) in the IR regime. Thus, we do not
need to improve it here to make it integrable and we
note that the singular contributions in the other
kernels are expected to dominate the evolution effects in any case. We do not exclude improving it for the best precision~\cite{elswh} and we
return to this
point presently.\par

This brings us then to $P_{GG}(z)$. Its lowest order form is
\begin{equation}
P_{GG}(z)= 2C_{G}(\frac{1-z}{z}+\frac{z}{1-z}+z(1-z))
\label{dglap17}
\end{equation} 
which again exhibits unintegrable IR singularities at both $z=1$ and $z=0$.
(Here, $C_G$ is the gluon quadratic Casimir invariant, so that it is just 
$N_c=3$.) If we repeat the QCD exponentiation calculation carried-out above
by using the color representation for the gluon rather than that for the
quarks, i.e., if we apply the exponentiation analysis in the Appendix to
the squared amplitude for the process $G\rightarrow G(z)+G(1-z)$, we get the exponentiated un-normalized result{\small
\begin{equation}
P_{GG}(z)= 2C_G F_{YFS}(\gamma_G)e^{\frac{1}{2}\delta_G}\left(\frac{1-z}{z}z^{\gamma_G}+\frac{z}{1-z}(1-z)^{\gamma_G}+\frac{1}{2}(z^{1+\gamma_G}(1-z)+z(1-z)^{1+\gamma_G})\right)
\label{dglap18}
\end{equation}}
wherein we obtain the $\gamma_G$ and $\delta_G$ from the expressions for
$\gamma_q$ and $\delta_q$ by the substitution $C_F\rightarrow C_G$:
\begin{align}
\gamma_G &= C_G\frac{\alpha_s}{\pi}t=\frac{4C_G}{\beta_0}\\
\delta_G&=\frac{\gamma_G}{2}+\frac{\alpha_sC_G}{\pi}(\frac{\pi^2}{3}-\frac{1}{2}).
\label{dglap19}
\end{align}
We see again that exponentiation has again made the singularities at 
$z=1$ and $z=0$ integrable.\par

To normalize $P_{GG}$, we take into account the virtual corrections such that
the gluon momentum sum rule  
\begin{equation}
\int_{0}^{1}dz z \left(2n_f P_{qG}(z)+P_{GG}(z)\right) = 0
\label{dglap20}
\end{equation}
is satisfied.
This gives us finally the IR-improved result
\begin{equation}
\begin{split}
P_{GG}(z)&= 2C_G F_{YFS}(\gamma_G)e^{\frac{1}{2}\delta_G}\{ \frac{1-z}{z}z^{\gamma_G}+\frac{z}{1-z}(1-z)^{\gamma_G}\\
&\qquad +\frac{1}{2}(z^{1+\gamma_G}(1-z)+z(1-z)^{1+\gamma_G}) - f_G(\gamma_G) \delta(1-z)\}
\end{split}
\label{dglap21}
\end{equation}
where for $f_G(\gamma_G)$ we get
\begin{equation}
\begin{split}
f_G(\gamma_G)&=\frac{n_f}{6C_GF_{YFS}(\gamma_G)}{e^{-\frac{1}{2}\delta_G}}+
\frac{2}{\gamma_G(1+\gamma_G)(2+\gamma_G)}+\frac{1}{(1+\gamma_G)(2+\gamma_G)}\\
&\qquad +\frac{1}{2(3+\gamma_G)(4+\gamma_G)}+\frac{1}{(2+\gamma_G)(3+\gamma_G)(4+\gamma_G)}.
\end{split}
\end{equation}
It is these improved results in eqs.(\ref{dglap11},\ref{dglap13},\ref{dglap21})
for $P_{qq}(z)$, $P_{Gq}(z)$ and $P_{GG}(z)$ that we 
use together with the standard result in (\ref{dglap17}) for
$P_{qG}(z)$ as the IR-improved DGLAP-CS theory.\par

For clarity we summarize at this point 
the new IR-improved kernel set as follows:
\begin{align}
P_{qq}^{exp}(z)&= C_F F_{YFS}(\gamma_q)e^{\frac{1}{2}\delta_q}\left[\frac{1+z^2}{1-z}(1-z)^{\gamma_q} -f_q(\gamma_q)\delta(1-z)\right],\\
P_{Gq}^{exp}(z)&= C_F F_{YFS}(\gamma_q)e^{\frac{1}{2}\delta_q}\frac{1+(1-z)^2}{z} z^{\gamma_q},\\
P_{GG}^{exp}(z)&= 2C_G F_{YFS}(\gamma_G)e^{\frac{1}{2}\delta_G}\{ \frac{1-z}{z}z^{\gamma_G}+\frac{z}{1-z}(1-z)^{\gamma_G}\nonumber\\
&\qquad +\frac{1}{2}(z^{1+\gamma_G}(1-z)+z(1-z)^{1+\gamma_G}) - f_G(\gamma_G) \delta(1-z)\},\\
P_{qG}(z)&=\frac{1}{2}(z^2+(1-z)^2),
\label{dglap22}
\end{align}
where we have introduced the superscript $exp$ to denote the exponentiated
results henceforth.\par

Returning now to the improvement of $P_{qG}(z)$, let us record it as well
for the sake of completeness and of providing better precision. 
Applying eq.(\ref{subp15}) to the
process $G\rightarrow q+\bar{q}$, we get the exponentiated result
\begin{equation}
P_{qG}^{exp}(z)= F_{YFS}(\gamma_G)e^{\frac{1}{2}\delta_G}\frac{1}{2}\{ z^2(1-z)^{\gamma_G}+(1-z)^2z^{\gamma_G}\}.
\label{dglap221}
\end{equation}
The gluon momentum sum rule then gives the new normalization constant for the
$P_{GG}^{exp}$ via the result
\begin{equation}
\begin{split}
\bar{f}_G(\gamma_G)&=\frac{n_f}{C_G}\frac{1}{(1+\gamma_G)(2+\gamma_G)(3+\gamma_G)}+
\frac{2}{\gamma_G(1+\gamma_G)(2+\gamma_G)}+\frac{1}{(1+\gamma_G)(2+\gamma_G)}\\
&\qquad +\frac{1}{2(3+\gamma_G)(4+\gamma_G)}+\frac{1}{(2+\gamma_G)(3+\gamma_G)(4+\gamma_G)}.
\end{split}
\label{dglap222}
\end{equation} 
The constant $\bar{f}_G$ should be substituted for $f_G$ in $P_{GG}^{exp}$
whenever the exponentiated result in (\ref{dglap221}) is used.
These results (\ref{dglap22}),~(\ref{dglap221}), and ~(\ref{dglap222})
are our new improved DGLAP-CS kernel set, with the option
exponentiating $P_{qG}$ as well.\par

In the discussion so far, we have used the lowest order DGLAP-CS
kernel set to illustrate how important the resummation which we present
here can be. In the literature~\cite{mvermn,mvovermn}, there are now exact results up to ${\cal O}(\alpha_s^3)$ for the DGLAP-CS kernels. The question naturally arises as to the relationship of our work to these fixed-order exact results.
We stress first that we are presenting an improvement of the fixed-order
results such that the singular pieces of the any exact fixed-order
result, i.e., the $\frac{1}{(1-z)_+}$ parts, are exponentiated so that they are
replaced with integrable functions proportional to $(1-z)^{\gamma-1}$ 
with $\gamma$ positive as we have illustrated above. Since the series
of logs which we resum to accomplish this has the structure
$\alpha_s^\ell t^\ell\ln^n(1-z),~\ell\ge n$
these terms are not already present
in the results in Refs.~\cite{mvermn,mvovermn}. As we use the formula
in eq.(\ref{subp15}), there will be no double counting if we
implement our IR-improvement of the exact fixed-order results in
Refs.~\cite{mvermn,mvovermn}. The detailed discussion of the
application of our theory to the results in Refs.~\cite{mvermn,mvovermn}
will appear elsewhere~\cite{elswh}. For reference, we note that the
higher order kernel corrections in Refs.~\cite{mvermn,mvovermn} are
perturbatively related to the leading order kernels, so one can expect that
the size
of the exponentiation effects illustrated above and below 
will only be perturbatively
modified by the higher order kernel corrections, leaving the same
qualitative behavior in general.
\par

In the interest of specificness, let us illustrate the IR-improvement
of $P_{qq}$ when calculated to three loops using the results
in Refs.~\cite{mvermn,mvovermn}. Considering the non-singlet case
for definiteness (a similar analysis holds for the singlet case)
we write in the notation of the latter references
\begin{equation}
P_{ns}^{+}=P_{qq}^v+P_{q\bar{q}}^v\equiv \sum_{n=0}^\infty(\frac{\alpha_s}{4\pi})^{n+1}P_{ns}^{(n)+}
\label{vermn1}
\end{equation}
where at order ${\cal O}(\alpha_s)$ we have 
\begin{equation}
P_{ns}^{(0)+}(z)=2C_F\{\frac{1+z^2}{(1-z)_+}+\frac{3}{2}\delta(1-z)\}
\label{vermn2}
\end{equation}
which shows that $P_{ns}^{(0)+}(z)$ agrees with the unexponentiated result in
(\ref{dglap7}) for $P_{qq}$ except for an overall factor of 2.
We use this latter identification to connect our work with that
in Refs.~\cite{mvermn,mvovermn} in the standard methodology.
In Refs.~\cite{mvermn,mvovermn}, exact results are given for
$P_{ns}^{(1)+}(z)$, and in Refs.~\cite{mvovermn} exact results are given
for $P_{ns}^{(2)+}(z)$.
When we apply the result in (\ref{subp15})
to the squared amplitudes for the processes $q\rightarrow q+X$, $\bar{q}\rightarrow q+X'$, we get the exponentiated result
\begin{equation}
\begin{split}
P_{ns}^{+,exp}(z)&=(\frac{\alpha_s}{4\pi}) 2P_{qq}^{exp}(z)+F_{YFS}(\gamma_q)e^{\frac{1}{2}\delta_q}\big{[}(\frac{\alpha_s}{4\pi})^2\{(1-z)^{\gamma_q}\bar{P}_{ns}^{(1)+}(z)+\bar{B}_2\delta(1-z)\}\\
&\qquad +(\frac{\alpha_s}{4\pi})^3\{(1-z)^{\gamma_q}\bar{P}_{ns}^{(2)+}(z)+\bar{B}_3\delta(1-z)\}\big{]}
\end{split}
\label{vermn3}
\end{equation} 
where $P_{qq}^{exp}(z)$ is given in (\ref{dglap22}) and the resummed residuals 
$\bar{P}_{ns}^{(i)+}$,~$i=1,2$ are related to the exact results for
$P_{ns}^{(i)+}$,~$i=1,2$, as follows:
\begin{equation}
\bar{P}_{ns}^{(i)+}(z)=P_{ns}^{(i)+}(z)-B_{1+i}\delta(1-z)+\Delta_{ns}^{(i)+}(z)
\label{vermn4}
\end{equation}
where 
\begin{equation}
\begin{split}
\Delta_{ns}^{(1)+}(z)&=-4C_F\pi\delta_1\{\frac{1+z^2}{1-z}-f_q\delta(1-z)\}\\
\Delta_{ns}^{(2)+}(z)&=-4C_F(\pi\delta_1)^2\{\frac{1+z^2}{1-z}-f_q\delta(1-z)\}\\
 &\qquad \qquad -2\pi\delta_1\bar{P}_{ns}^{(1)+}(z)
\end{split}
\label{vermn5}
\end{equation}
and
\begin{equation}
\begin{split}
\bar{B}_2&=B_2+4C_F\pi\delta_1f_q\\
\bar{B}_3&=B_3+4C_F(\pi\delta_1)^2f_q-2\pi\delta_1\bar{B}_2.
\end{split}
\label{vermn6}
\end{equation}
Here, the constants $B_i,~i=2,3$ are given by the results in
Refs.~\cite{mvermn,mvovermn} as
\begin{equation}
\begin{split}
B_2&=4C_GC_F(\frac{17}{24}+\frac{11}{3}\zeta_2-3\zeta_3)-4C_Fn_f(\frac{1}{12}+\frac{2}{3}\zeta_2)+4C_F^2(\frac{3}{8}-3\zeta_2+6\zeta_3)\\
B_3&=16C_GC_Fn_f(\frac{5}{4}-\frac{167}{54}\zeta_2+\frac{1}{20}\zeta_2^2+\frac{25}{18}\zeta_3)\\
&\qquad +16C_GC_F^2(\frac{151}{64}+\zeta_2\zeta_3-\frac{205}{24}\zeta_2-\frac{247}{60}\zeta_2^2+\frac{211}{12}\zeta_3+\frac{15}{2}\zeta_5)\\
&\qquad +16C_G^2C_F(-\frac{1657}{576}+\frac{281}{27}\zeta_2-\frac{1}{8}\zeta_2^2-\frac{97}{9}\zeta_3+\frac{5}{2}\zeta_5)\\
&\qquad +16C_Fn_F^2(-\frac{17}{144}+\frac{5}{27}\zeta_2-\frac{1}{9}\zeta_3)\\
&\qquad +16C_F^2n_F(-\frac{23}{16}+\frac{5}{12}\zeta_2+\frac{29}{30}\zeta_2^2-\frac{17}{6}\zeta_3)\\
&\qquad +16C_F^3(\frac{29}{32}-2\zeta_2\zeta_3+\frac{9}{8}\zeta_2+\frac{18}{5}\zeta_2^2+\frac{17}{4}\zeta_3-15\zeta_5),
\end{split}
\end{equation}
where $\zeta_n$ is the Riemann zeta function evaluated at argument $n$. 
The detailed phenomenological consequences of the fully exponentiated
2- and 3-loop DGLAP-CS kernel set will appear elsewhere~\cite{elswh}.
\par

\section{Phenomenological Effects of IR-Improvement}

Let us now look into the effects of IR-improvement 
on the moments of the structure
functions by discussing the corresponding effects on the moments
of the parton distributions. In this section we work with
the leading order results for definiteness of illustration. \par

We know that moments of the kernels determine the exponents in the logarithmic
variation~\cite{dglap,pgp,djgfw} of the moments of the quark distributions
and, thereby, of the moments of the structure functions themselves.
To wit, in the non-singlet case, we have 
\begin{equation}
\frac{dM^{NS}_n(t)}{dt}=\frac{\alpha_s(t)}{2\pi}A^{NS}_nM^{NS}_n(t)
\label{dglap23}
\end{equation}
where
\begin{equation}
M^{NS}_n(t)=\int^1_0dz z^{n-1}q^{NS}(z,t)
\label{dglap24}
\end{equation}
and the quantity $A^{NS}_n$ is given (using (\ref{dglap22})) by
\begin{align}
A^{NS}_n&=\int^1_0dz z^{n-1}P_{qq}^{exp}(z),\nonumber\\
        &= C_F F_{YFS}(\gamma_q)e^{\frac{1}{2}\delta_q}[B(n,\gamma_q)+B(n+2,\gamma_q)-f_q(\gamma_q)]
\label{dglap25}
\end{align}
where $B(x,y)$ is the beta function
given by \[B(x,y)=\Gamma(x)\Gamma(y)/\Gamma(x+y)\].
This should be compared to the un-IR-improved result~\cite{dglap,pgp,djgfw}:
\begin{equation}
A^{NS^o}_n\equiv C_F\left[-\frac{1}{2}+\frac{1}{n(n+1)}-2\sum_{j=2}^{n}\frac{1}{j}\right].
\label{dglap26}
\end{equation}
The asymptotic behavior for large $n$ is now very different, as
the IR-improved exponent approaches a constant, a multiple of $-f_q$,
as we would expect as $n\rightarrow \infty$ because
$\lim_{n\rightarrow \infty}z^{n-1} = 0$ for $0\le z<1$ 
whereas, as it is well-known,
the un-IR-improved result in (\ref{dglap26}) diverges as $-2C_F\ln n$
as $n\rightarrow \infty$. The two results are also different at finite
$n$: for $n=2$ we get, for example, for $\alpha_s\cong .118$~\cite{siggi},
\begin{equation}
A^{NS}_2 =
\begin{cases}
C_F(-1.33)&,~~\text{ un-IR-improved}\\
C_F(-0.966)&,~~\text{IR-improved}
\end{cases} 
\label{dglap27}
\end{equation}
so that the effects we have calculated are important for all $n$
in general. For completeness, we note that the solution to (\ref{dglap23})
is given by the standard methods as
\begin{equation}
\begin{split}
M^{NS}_n(t)&=M^{NS}_n(t_0)e^{\int_{t_0}^{t}dt'\frac{\alpha_s(t')}{2\pi}A^{NS}_n(t')}\\
&=M^{NS}_n(t_0)e^{\bar{a}_n[Ei(\frac{1}{2}\delta_1\alpha_s(t_0))-Ei(\frac{1}{2}\delta_1\alpha_s(t))]} \\
&\qquad \operatornamewithlimits{\Longrightarrow}_{\small t,t_0~\text{large with}~t>>t_0}M^{NS}_n(t_0)\left(\frac{\alpha_s(t_0)}{\alpha_s(t)}\right)^{\bar{a'}_n}
\end{split}
\label{dglap27a} 
\end{equation}
where $Ei(x)=\int_{-\infty}^xdre^r/r$ is the exponential integral function,
\begin{equation}
\begin{split}
\bar{a}_n&=\frac{2C_F}{\beta_0}F_{YFS}(\gamma_q)e^{\frac{\gamma_q}{4}}[B(n,\gamma_q)+B(n+2,\gamma_q)-f_q(\gamma_q)]\\
\bar{a'}_n&=\bar{a}_n\left(1+\frac{\delta_1}{2}\frac{(\alpha_s(t_0)-\alpha_s(t))}{\ln(\alpha_s(t_0)/\alpha_s(t))}\right)
\end{split}
\label{dglap27b}
\end{equation}
with
\[\delta_1=\frac{C_F}{\pi}\left(\frac{\pi^2}{3}-\frac{1}{2}\right)\].
We can compare with the un-IR-improved result in which the last line
in eq.(\ref{dglap27a}) holds exactly with $\bar{a'}_n=2A^{NS^o}_n/\beta_0$.
Phenomenologically, for $n=2$, taking $Q_0=2$GeV and evolving to $Q=100$GeV,
if we set $\Lambda_{QCD}\cong .2GeV$ and use $n_f=5$ for definiteness of illustration, we see from eqs. (\ref{dglap27a},\ref{dglap27b}) 
that we get a shift of the 
respective evolved NS moment by $\sim~5\%$, 
which is of some interest in view of the
expected HERA precision~\cite{carli}.\par
We stress that the size of the exponent $\gamma_q$ is what one would expect
from analogy with QED~\cite{yfs1}, where with $Q=100$ GeV we have the
analogous result $\gamma_e=(\alpha_{EM}/\pi)(\ln Q^2/m_e^2 -1)\cong 0.054$
whereas here, with $\alpha_s\cong .118$, so that it is about 10 times $\alpha_{EM}$, we get a value for $\gamma_q$ that is about 10 times $gamma_e$.
There are no data that currently contradict the values we find for our
$\gamma_j$,~$=q,G$. Evidently, from the exact result (\ref{expker1}) we can consistently improve the values of the $\gamma_j$ as needed using more and more of
the perturbative results for the functions $SUM_{IR}(QCD), \tilde S_{\rQCD},\tilde{\bar\beta}_n$ on the RHS accordingly, as we have noted.\par  

We give now the remaining elements of the leading anomalous dimension
matrix in its 'best' IR-improved form for completeness:
\begin{align}
A^{Gq}_n&=\int_0^1dz z^{n-1} P^{exp}_{Gq}(z)= C_F F_{YFS}(\gamma_q)e^{\frac{1}{2}\delta_q}
\left[\frac{1}{n+\gamma_q-1}+B(3,n+\gamma_q-1)\right],\\
A^{GG}_n&=\int_0^1dz z^{n-1}P^{exp}_{GG}(z)= 2C_G F_{YFS}(\gamma_G)e^{\frac{1}{2}\delta_G}\{ B(n+1,\gamma_G)+B(n+\gamma_G-1,2)\nonumber\\
&\qquad +\frac{1}{2}(B(n+1,\gamma_G+2)+B(n+\gamma_G+1,2))-\bar{f}(\gamma_G)\},\\
2n_fA^{qG}_n&=2n_f\int^1_0dz z^{n-1}P^{exp}_{qG}(z)=2T(F)F_{YFS}(\gamma_G)e^{\frac{1}{2}\delta_G}\left(B(n+2,1+\gamma_G)+B(n+\gamma_G,3)\right),
\label{dglap28}
\end{align}
where $T(F)=\frac{1}{2}n_f$. We 
note that the un-exponentiated value of the 
last result in eq.(\ref{dglap28}) is
a well-known one~\cite{dglap,pgp,djgfw}, $2T(F)\frac{2+n+n^2}{n(n+1)(n+2)}$,
and it would be used whenever we do not choose to exponentiate $P_{qG}$.
We will investigate the further implications of these IR-improved
results for LHC physics elsewhere~\cite{elswh}.\par

\section{Impact of IR-Improvement on the Standard Methodology}

In sum, we have used exact re-arrangement of the QCD Feynman
series to isolate and resum the leading IR contributions
to the physical processes that generate the evolution kernels
in DGLAP-CS theory. In this way, we have obviated the need to
employ artificial mathematical regularization of the attendant
IR singularities as the theory's higher order corrections
naturally tame these singularities. The resulting IR-improved 
anomalous dimension
matrix behaves more physically for large $n$ and receives significant
effects at finite $n$ from the exponentiation.\par

We in principle
can make contact with the moment-space resummation results
in Ref.~\cite{moment} but we stress that these results have necessarily been
obtained after making a Mellin transform of the mathematical
artifact which we address in this paper. Thus, the results
in Ref.~\cite{moment} do not in any way contradict the analysis 
in this paper.\par 

We note that the program of improvement of the hadron cross section
calculations for LHC physics advanced herein should be distinguished 
from the results in Refs.~\cite{cosopster,sterman,cattrent}. 
Indeed, recalling the
standard hadron cross section formula
\begin{equation}
\sigma =\sum_{i,j}\int dx_1dx_2F_i(x_1)F_j(x_2)\hat\sigma(x_1x_2s)
\label{sigtot}
\end{equation}
where $\{F_\ell(x)\}$ are the respective parton densities and
$\hat\sigma(x_1x_2s)$ is the respective reduced hard parton cross section,
the resummation results in Refs.~\cite{sterman,cattrent} address,
by summing the large logs in Mellin transform space, 
the $x_1x_2\rightarrow 1$ limit of $\hat\sigma(x_1x_2s)$
whereas the results above address the improvement, by resummation
in x-space, of the calculation of the parton densities $\{F_i(x)\}$ for all
values of x. Thus, the program of improvement presented above is entirely 
complementary to that in Refs.~\cite{sterman,cattrent} and 
both programs of improvement are useful for precision LHC physics.
The situation can be illustrated by comparing the 
results in Refs.~\cite{mochvv} with our results herein. The key
observation can already be made from eq.(2.1) in the latter paper,
wherein it is made manifest that the resummation carried out therein,
as an application of the methods in Refs.~\cite{sterman,cattrent}, is
a resummation for the large N-Mellin space limit of the Mellin transform
of the hard scattering coefficient function so that all of the IR effects
in the parton densities are not included in this resummation. What we
deal with here is however resummation of the IR effects in the kernels
which generate exactly these IR effects in these parton densities directly
in configuration space so that we work on a complementary aspect the
formula (\ref{sigtot}) and this we do directly in x-space rather than in
N-Mellin space. There is then no contradiction or repetition between
our results and those in Ref.~\cite{mochvv}. Refs.~\cite{sterman,cattrent}
have stressed that the IR singularities of the parton densities
are not addressed by their analysis. If we wanted to include
the results in Refs.~\cite{sterman,cattrent}, as we have discussed in Refs.~\cite{qcdexp}, we would make an inverse Mellin transform of their results and
apply them to our resummation calculus for the hard cross section
$\hat\sigma(x_1x_2s)$ as it is illustrated in the Appendix; by construction,
this would not affect the IR singularities for the evolution
of the initial parton densities analyzed in this paper. 
\par
Similarly, the analysis of the $P_T$ distribution resummation in Ref.~\cite{cosopster} deals with the hard scattering coefficient response to soft-gluon 
emission. The initial parton densities are input to this analysis. 
Evidently, as we work with the x-space initial parton density evolution 
in which any $P_T$ has been
integrated out, the analysis in Ref.~\cite{cosopster} is entirely complementary
to our work in this paper.~\par

Finally, we address the issue of the relationship between the re-arrangement 
that we have made of the exact leading-logs in the QCD perturbation theory
and the usual treatment in the non-exponentiated DGLAP-CS theory. If one
expands out the exponentiated kernels, using the distribution identity
\begin{equation}
(1-z)^{a-1}=\frac{1}{a}\delta(1-z)+\frac{1}{(1-z)_+}+\sum_{j=1}^{\infty}\frac{a^j}{j!}\left[\frac{\ln^j(1-z)}{1-z}\right]_+,
\label{dist1}
\end{equation} 
one can see that for example $P_{qq}$ and $P^{exp}_{qq}$ agree 
to leading order, so that the leading log series which they 
generate for the respective NS
parton distributions 
also agree through leading order in $\frac{\alpha_s}{\pi}L$
where $L$ is the respective big log in momentum-space. At higher orders then,
we have a different result for the $\{F_i\}$, let us denote them 
by $\{{F'}_i\}$, and a different result for the reduced cross section, 
let us denote it by $\hat\sigma'$, such that we get the 
same perturbative QCD cross section,
\begin{equation}
\begin{split}
\sigma &=\sum_{i,j}\int dx_1dx_2F_i(x_1)F_j(x_2)\hat\sigma(x_1x_2s)\cr
       &=\sum_{i,j}\int dx_1dx_2{F'}_i(x_1){F'}_j(x_2)\hat\sigma'(x_1x_2s)
\end{split}
\end{equation}
order by order in perturbation theory. The exponentiated kernels are
used to factorize the mass singularities from the unfactorized reduced
cross section and this generates $\hat\sigma'$ instead of the usual
$\hat\sigma$ whose factorized form is generated using the usual DGLAP-CS kernels.
For example, from (\ref{dglap7}), (\ref{dglap22}) and (\ref{dist1}),
we see that, 
starting at order ${\cal O}(\alpha_s^2)$,
when one factorizes the mass singularities using $P_{qq}^{exp}$ one makes
a different subtraction from the respective fixed-order unfactorized
hard cross section compared to what one would subtract if the factorization
were done with the unexponentiated $P_{qq}$, resulting in a different reduced
cross section, $\hat\sigma'$; but, 
the convloution of this different 
reduced cross section with the respective $\{F_i'\}$
that result from solving the DGLAP-CS equation with the IR-improved kernels
would then give the same result for the hadron-hadron scattering cross section
as one would get with the un-improved set $\{F_i,~\hat\sigma\}$, order by
order in perturbation theory.\par

The entirely analogous statement holds for the structure functions
in deep inelastic scattering. They are represented as a 
sum over products of the parton distributions and the attendant
hard scattering cross sections. Again, if we use exponentiated kernels
to generate the respective parton distributions, we use these kernels
to isolate the respective factorized hard scattering cross sections,
giving the same perturbative QCD prediction order by order in perturbation
theory for the respective deep inelastic structure functions.  
\par

We thus have the same leading log series for $\sigma$ as does the usual
calculation with un-exponentiated DGLAP-CS kernels. We have an important
advantage: the lack of +-functions in the generation of the 
configuration space functions
$\{{F'}_i,~\hat\sigma'\}$ means that these functions lend themselves 
more readily to Monte Carlo realization to arbitrarily soft radiative
effects, both for the generation of the parton shower associated
to the $\{{F'}_i\}$ and for the attendant remaining 
radiative effects in $\hat\sigma'$.
\par
We stress that, as on can see from our analysis above, unlike the
standard $\overline{MS}$ unexponentiated kernels, our exponentiated kernels
contain powers of the product $\alpha_sL$, as these describe the large
IR effects which we resum. These effects are then beyond the standard
Wilson expansion, as they cause non-trivial modification of the moments
of the exponentiated kernels relative to the moments
of the $\overline{MS}$ kernels and it is the latter which
correspond to the respective anomalous dimensions of the operators
in Wilson's expansion when this expansion exists. We exchange the naive
connection between the moments of the kernels and these anomalous
dimensions for the improved IR behavior of the exponentiated kernels.
We do this with an eye toward MC methods.\par

The use of the new IR-improved kernels may seem as though one is
over counting effects; after all, it is well-known~\cite{exactsoln} 
that the response
of the DGLAP-CS equation for a delta-function at $z=1$ for an initial
value $t=t_0$ in the NS case is indeed a solution which exhibits
exactly $(1-z)^{\gamma_q-1}$ type behavior for the respective
distribution that we have found for our NS exponentiated kernel.
We are not over-counting this effect. For, consider the sequential
application of the kernel $P_{qq}^{exp}(z)$ so that, first, a quark, $q(1)$,
splits via its action to a quark $q(z)$ and gluons $\{G_i(\xi_i)\}$,
with $\sum_i \xi_i=1-z$ in the standard notation, and, second, the latter
quark, $q(z)$, splits via the second application of $P_{qq}^{exp}(z')$
to a quark $q(zz')$ and gluons $\{G_j(\xi'_j)\}$, with $\sum_j\xi'_j=(1-z')z$.
In the first action, the quantum transition was for $q(1)\rightarrow q(z)$
on the quark Hilbert space while in the second it was 
for $q(z)\rightarrow q(zz')$, a completely different quantum transition
on the quark Hilbert space. The two transitions are completely
independent in the leading log and using $P_{qq}^{exp}$ to effect them
does not over-count anything; it simply makes the description of 
each transition more accurate.  

\section{Conclusions}

We have developed a new approach to precision QCD predictions for
high energy colliding beam physics scenarios such as that afforded us
by the impending turn-on of the LHC. We do not contradict any aspect
of the traditional approach. We gain the advantages of improved IR
behavior for the respective kernels, parton distributions 
and reduced cross sections,
which should facilitate realization by multiple gluon MC methods.
Such realizations for LHC physics will appear elsewhere.~\cite{elswh}.

\section*{Acknowledgments}
We thank Prof. S. Jadach for useful discussions and Prof. W. Hollik
for the kind hospitality of the Max-Planck-Institut, Munich, wherein
a part of this work was completed.

\section*{Appendix }

In this Appendix I, we present the new QCD exponentiation theory
which has been
developed in Refs.~\cite{qcdexp,delaney} as it is not generally
familiar. The goal is to make the current paper self-contained.\par

For definiteness,
we will use the process in Fig.~2, $\bar Q'(p_1) Q(q_1)\rightarrow \bar Q'''(p_2) Q''(q_2) +
G_1(k_1)\cdots G_n(k_n)$, as the proto-typical process, 
\begin{figure}
\begin{center}
\setlength{\unitlength}{1mm}
\begin{picture}(160,80)
\put(40, -8){\makebox(0,0)[lb]{
\epsfig{file=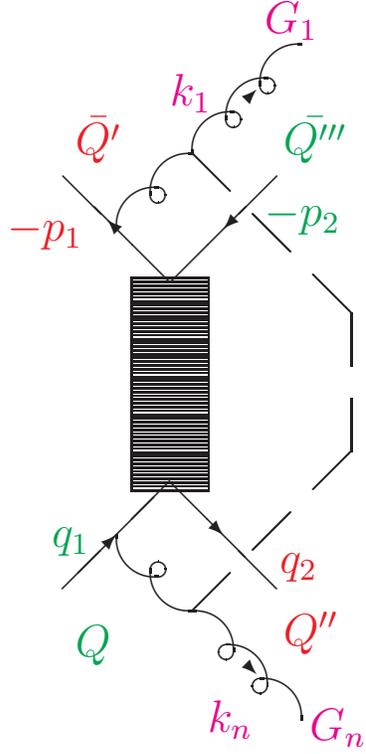,width=50mm}
}}
\end{picture}
\end{center}
\caption{\baselineskip=7mm     The process $\bar{Q'}
    Q      \rightarrow        \bar{Q'''}
                 +      Q''    + n(G)$.
The four--momenta are indicated in the standard manner: $q_1$ 
is the
four--momentum of the incoming $Q$, $q_2$ 
is the four--momentum of the outgoing $Q''$, etc.,
and $Q = u,d,s,c,b,G.$}
\label{figproto.1}
\end{figure}
\noindent
where we have written the
kinematics as it is illustrated in the figure. This process,
which dominates processes such as t\=t production at FNAL, contains all of the 
theoretical issues that we must face at the parton level to establish
, as an extension of the original 
ideas of Yennie, Frautschi and Suura (YFS)~\cite{yfs}, 
QCD soft exponentiation by MC methods -- applicability to other related
processes will be immediate. For reference, let us also 
note that, in what follows, we use the GPS conventions of 
Ref.~\cite{gps} for spinors $\{u,v,\umf\}$ and the attendant
photon and gluon polarization vectors that follow therefrom:
\begin{equation}
\label{gmapol}
  (\epsilon^\mu_\sigma(\beta))^*
     ={\bar{u}_\sigma(k) \gamma^\mu u_\sigma(\beta)
       \over \sqrt{2}\; \bar{u}_{-\sigma}(k) u_\sigma(\beta)},\quad
  (\epsilon^\mu_\sigma(\zeta))^*
     ={\bar{u}_\sigma(k) \gamma^\mu \umf_\sigma(\zeta)
       \over \sqrt{2}\; \bar{u}_{-\sigma}(k) \umf_\sigma(\zeta)},
\end{equation}
with $\beta^2=0$ and $\zeta$ defined in Ref.~\cite{gps}, 
so that all phase information is strictly known in our amplitudes.
This means that, although we shall use the older EEX realization of YFS
MC exponentiation as defined in Ref.~\cite{ceex:2001}, the realization
of our results via the the newer CEEX realization of YFS exponentiation
in Ref.~\cite{ceex:2001} is also possible and is in progress~\cite{elswh}.

Specifically, the authors in Refs.~\cite{delaney}
have analyzed how in the special case of Born level color exchange
one applies the YFS theory to QCD by extending
the respective YFS
IR singularity analysis to QCD to all orders in $\alpha_s$.
Here, unlike what was emphasized in Refs.~\cite{delaney},
we focus on the YFS theory as a general re-arrangement of
renormalized perturbation theory based on its IR
behavior, just as the renormalization group
is a general property of renormalized perturbation theory based on its
UV(ultra-violet) behavior.
We will thus keep our arguments
entirely general from the outset, so that it will be immediate
that our result applies to any renormalized perturbation theory
in which the cross section under study is finite.\par

Let the amplitude for the emission of $n$ real gluons in our proto-typical
subprocess, 
$Q^\alpha + {\bar {Q'}}^{\bar\alpha} \rightarrow {Q''}^\gamma
{\bar {Q'''}}^{\bar\gamma} + n(G)$, where $\alpha, {\bar\alpha}, \gamma$, and
${\bar\gamma}$ are color indices, be
represented by
\begin{equation}
{\cal M}^{(n)\alpha{\bar\alpha}}_{\gamma{\bar\gamma}} 
 = \sum_{\ell}M^{(n)\alpha{\bar\alpha}}_{\gamma{\bar\gamma}\ell},
\label{subp1}
\end{equation}
$M^{(n)}_{\ell}$ is the contribution to 
${\cal M}^{(n)}$
from Feynman diagrams with ${\ell}$ virtual loops.
Symmetrization yields
\begin{eqnarray}
   M^{(n)}_\ell = {1\over {\ell!}}\int\prod_{j=1}^{\ell}{d^4k_j\over{(2\pi)^4(
 k_j^2-\lambda^2+i\epsilon)}}\rho^{(n)}_\ell(k_1,\cdots,k_\ell),
\label{subp2}
\end{eqnarray}
where this last equation defines $\rho^{(n)}_\ell$ as a symmetric
function of its arguments arguments $k_1,...,k_{\ell}$.
$\lambda$ will be our infrared gluon regulator mass for IR singularities;
n-dimensional regularization of the 't Hooft-Veltman~\cite{tHvelt} type is also
possible as we shall see.

We now define the virtual IR emission factor $S_{QCD}(k)$ for a gluon
of 4-momentum $k$, for the $k\rightarrow 0$ regime of the
respective 4-dimensional loop integration as in (\ref{subp2}), such that
\begin{equation}
\lim_{k\rightarrow 0}k^2\left( \rho^{(n)\alpha{\bar\alpha}}_{\gamma{\bar\gamma}1}(k)|_{\text{leading Casimir contribution}}
 -S_{QCD}(k)\rho^{(n)\alpha{\bar\alpha}}_{\gamma{\bar\gamma}0}\right) = 0,
\end{equation}
where we have now introduced the restriction to the leading
color Casimir terms at one-loop\footnote{These correspond with
maximally non-Abelian terms in Ref.~\cite{gatherall} but computed
exactly rather than in the eikonal approximation.} 
so that in the expression for the respective one-loop correction 
$\rho^{(n)}_1$ and in that for  
for $S_{QCD}(k)$ given in Refs.~\cite{delaney}, only the
terms proportional to $C_F$ should be retained here as
we focus on the f\=f$\rightarrow$f\=f case, where f denotes a fermion.
(Henceforth, when we refer to $k\rightarrow 0$ gluons we are always referring
for virtual gluons to the corresponding regime of the 4-dimensional
loop integration in the computation of $ M^{(n)}_\ell$.)

In Ref.~\cite{delaney},
the respective authors have calculated $S_{QCD}(k)$
using the running quark masses to regulate its collinear
mass singularities, for example;  n-dimensional regularization of
the 't Hooft-Veltman type is also possible for
these mass singularities and we will also illustrate this presently.

We stress that $S_{QCD}(k)$
has a freedom in it corresponding to the fact that any
function $\Delta S_{QCD}(k)$ which has the property that
$\lim_{k\rightarrow 0} k^2\Delta S_{QCD}(k)\rho^{(n)}_0=0$
may be added to it.

Since the virtual gluons in $\rho^{(n)}_{\ell}$
are all on equal footing by the symmetry of this function,
if we look at gluon $\ell$, for example, we may write
, for $k_{\ell} \rightarrow (0,0,0,0)\equiv O$ while the
remaining $k_i$ are fixed away from $O$, the representation
\begin{equation}
 \rho^{(n)}_\ell = S_{QCD}(k_\ell)*\rho^{(n)}_{\ell-1}(k_1,\cdots,k_{\ell-1})
 +\beta^1_\ell(k_1,\cdots,k_{\ell-1};k_\ell)
\label{subp3}
\end{equation}
where the residual amplitude $\beta^1_\ell(k_1,\cdots,k_{\ell-1};k_\ell)$
will now be taken as defined by this last equation. It has two nice 
properties:
\begin{itemize} 
\item it is symmetric in its first $\ell - 1$ 
arguments 
\item the
IR singularities for gluon $\ell$ {\it that are contained in 
$S_{QCD}(k_\ell)$} are no longer contained in it. 
\end{itemize}

We do not at this point
discuss the extent to which there are any further remaining IR singularities
for gluon $\ell$ in $\beta^1_\ell(k_1,\cdots,k_{\ell-1};k_\ell)$.
In an Abelian gauge theory like QED, as has been shown by Yennie, Frautschi
and Suura in Ref.~\cite{yfs},
there would not be any further
such singularities; for a non-Abelian gauge theory like QCD, this point
requires further discussion and 
we will come back to this point presently.

We rather now stress that
if we apply the representation (\ref{subp3}) again we may write
{\small
\begin{eqnarray}
   \rho^{(n)}_\ell = S_{QCD}(k_\ell)S_{QCD}(k_{\ell-1})* \rho^{(n)}_{\ell-2}(k_1,\cdots,
 k_{\ell-2})\nonumber \\
 + S_{QCD}(k_\ell)\beta^1_{\ell-1}(k_1,\cdots,k_{\ell-2};k_{\ell-1})
\nonumber \\
+ S_{QCD}(k_{\ell-1})\beta^1_{\ell-1}(k_1,\cdots,k_{\ell-2};k_\ell)\nonumber \\
+ \beta^2_\ell(k_1,\cdots,k_{\ell-2};k_{\ell-1},k_\ell),
\label{subp4}
\end{eqnarray} 
where this last equation serves to define the function
$\beta^2_\ell(k_1,\cdots,k_{\ell-2};k_{\ell-1},k_\ell)$. It has two nice
properties:
\begin{itemize} 
\item it is symmetric in its first $\ell -2$ arguments
and in its last two arguments $k_{\ell-1},k_\ell$ 
\item the infrared
singularities for gluons $\ell-1$ and $\ell$ that are contained in
$S_{QCD}(k_{\ell-1})$ and $S_{QCD}(k_\ell)$ are no longer contained in
it. 
\end{itemize}

Continuing in this way, with repeated application of (\ref{subp3}),
we get finally the rigorous, exact rearrangement of the 
contributions to $\rho^{(n)}_\ell$ as
\begin{eqnarray}
\rho^{(n)}_\ell = S_{QCD}(k_1)\cdots S_{QCD}(k_\ell)\beta^0_0+\sum_{i=1}^\ell\prod_{j\neq i}
S_{QCD}(k_j)\beta^1_1(k_i) +\cdots \nonumber\\
+\beta^\ell_\ell(k_1,\cdots,k_\ell),
\label{subp5}
\end{eqnarray}
where the virtual gluon residuals  $\beta^i_i(k'_1,\cdots,k'_i)$
have two nice properties:
\begin{itemize} 
\item they are symmetric functions of their
arguments 
\item they do not contain any of the IR singularities which
are contained in the product 
$S_{QCD}(k'_1)\cdots S_{QCD}(k'_i)$.
\end{itemize}

Henceforth, we denote  
$\beta^i_i$ as the function $\beta_i$ for reasons of pedagogy.
We can not stress too much that (\ref{subp5}) is an {\it exact}
rearrangement of the contributions of the Feynman diagrams which
contribute to $\rho^{(n)}_\ell$; it involves no approximations.
Here also we note that the
question of the absolute convergence of these Feynman diagrams
from the standpoint of constructive field theory remains open as usual. Yennie,
Frautschi and Suura~\cite{yfs} have already stressed that Feynman diagrammatic
perturbation theory is non-rigorous from this standpoint.
What we do claim is that the relationship between the YFS expansion and the
usual perturbative Feynman diagrammatic expansion is itself rigorous
even though neither of the two expansions themselves is rigorous.
\par

Introducing (\ref{subp5}) into (\ref{subp1}) yields
a representation similar to that of YFS, and we
will call it a ``YFS representation'', 
\begin{equation}
{\cal M}^{(n)} = e^{\alpha_sB_{QCD}}\sum_{j=0}^\infty{\sf m}^{(n)}_j,
\label{yfsrepv}
\end{equation}
where we have defined
\begin{equation}
\alpha_s(Q)B_{QCD} = \int {d^4k\over (k^2-\lambda^2+i\epsilon)}S_{QCD}(k)
\label{vbfn}
\end{equation}
and
\begin{equation}
 {\sf m}^{(n)}_j = {1\over {j!}}\int\prod_{i=1}^j{d^4k_i\over k_i^2-\lambda^2+i\epsilon}
       \beta_j(k_1,\cdots,k_j). 
\label{irfreev}
\end{equation}
We say that (\ref{yfsrepv}) is similar to the respective result
of Yennie, Frautschi and Suura in Ref.~\cite{yfs}
and is not identical to it because we have not
proved that the functions $\beta_i(k_1,...,k_i)$ are completely
free of virtual IR singularities. What have shown is that they do not
contain the IR singularities in the product 
$S_{QCD}(k_1)\cdots S_{QCD}(k_i)$ so that
${\sf m}^{(n)}_j$ does not contain the virtual IR divergences
generated by this product when it is integrated over the respective
4j-dimensional j-virtual gluon phase space. In an Abelian gauge theory,
there are no other possible virtual IR divergences; in the non-Abelian
gauge theory that we treat here, such additional IR divergences
are possible and are expected; but, the 
result (\ref{yfsrepv}) does have an improved
IR divergence structure over (\ref{subp1}) in that all of the
IR singularities associated with $S_{QCD}(k)$ are explicitly
removed from the sum over the virtual IR improved loop contributions
${\sf m}^{(n)}_j$ to all orders in $\alpha_s(Q)$.\par

Turning now to the analogous rearrangement of the real IR singularities in
the differential cross section associated with the ${\cal M}^{(n)}$,
we first note that we may write this cross section as follows
according to the standard methods
\begin{eqnarray}
  d\hat\sigma^n = {e^{2\alpha_sReB_{QCD}}\over {n !}}\int\prod_{m=1}^n
{d^3k_m\over (k_m^2+\lambda^2)^{1/2}}\delta(p_1+q_1-p_2-q_2-\sum_{i=1}^nk_i)
\nonumber\\       
\bar\rho^{(n)}(p_1,q_1,p_2,q_2,k_1,\cdots,k_n)
{d^3p_2d^3q_2\over p^0_2 q^0_2},
\label{diff1}
\end{eqnarray}
where we have defined
\begin{equation}
\bar\rho^{(n)}(p_1,q_1,p_2,q_2,k_1,\cdots,k_n)=
\sum_{color,spin} \|\sum_{j=0}^\infty{\sf m}^{(n)}_j\|^2
\label{diff2}
\end{equation}}
in the incoming Q\=Q' cms system
and we have absorbed the remaining kinematical factors for
the initial state flux, spin and color averages into the
normalization of the amplitudes ${\cal M}^{(n)}$ for reasons of
pedagogy so that the $\bar\rho^{(n)}$ are averaged over initial spins
and colors and summed over final spins and colors.
We now proceed in complete analogy with the discussion
of $\rho^{(n)}_\ell$ above. \par

Specifically, 
for the functions $\bar\rho^{(n)}(p_1,q_1,p_2,q_2,k_1,\cdots,k_n)
\equiv \bar\rho^{(n)}(k_1,\cdots,k_n)$ which are symmetric functions
of their arguments $k_1,\cdots,k_n$, we define first, for $n=1$, 
\begin{equation}
\lim_{|\vec{k}|\rightarrow 0}\vec{k}^2\left(\bar\rho^{(1)}(k)|_{\text{leading Casimir contribution}}
-\tilde S_{QCD}(k)\bar\rho^{(0)}\right) = 0, 
\label{realS}
\end{equation}
where the real infrared function $\tilde S_{QCD}(k)$ is rigorously
defined by this last equation and is explicitly computed in
Refs.~\cite{delaney}, wherein we retain here only
the terms proportional to $C_F$ 
from the result in Ref.~\cite{delaney}
; like its virtual counterpart $S_{QCD}(k)$
it has a freedom in it in that any function $\Delta\tilde S_{QCD}(k)$
with the property that 
$\lim_{|\vec{k}|\rightarrow 0}\vec{k}^2\Delta\tilde S_{QCD}(k)=0$ 
may be added to it without affecting the defining relation (\ref{realS}).
 
We can again repeat the 
analogous arguments of Ref.~\cite{yfs}, following the
corresponding steps in (\ref{subp3})-(\ref{irfreev}) 
above for $S_{QCD}$ to get 
the ``YFS-like'' result {\small 
\begin{equation}
\begin{split}
d\hat\sigma_{\rm exp}&= \sum_n d\hat\sigma^n \\
         &=e^{\rm SUM_{IR}(QCD)}\sum_{n=0}^\infty\int\prod_{j=1}^n{d^3
k_j\over k^0_j}\int{d^4y\over(2\pi)^4}e^{iy\cdot(p_1+q_1-p_2-q_2-\sum k_j)+
D_\rQCD}\\
&*\bar\beta_n(k_1,\ldots,k_n){d^3p_2\over p_2^{\,0}}{d^3q_2\over
q_2^{\,0}}
\end{split}
\label{subp10}
\end{equation}
with 
\[ {SUM}_{IR}(QCD)=2\alpha_s ReB_{QCD}+2\alpha_s\tilde B_{QCD}(\Kmax),\]
\[ 2\alpha_s\tilde B_{QCD}(\Kmax)=\int{d^3k\over k^0}\tilde S_\rQCD(k)
\theta(\Kmax-k),\]
 \begin{equation} D_\rQCD=\int{d^3k\over k}\tilde S_\rQCD(k)
\left[e^{-iy\cdot k}-\theta(\Kmax-k)\right],\label{subp11}\end{equation}
\[{1\over 2}\bar\beta_0=d\sigma^{\rm(1-loop)}-2\alpha_s{\rm Re}B_\rQCD d\sigma_B,\]
\begin{equation}{1\over 2}\bar\beta_1=d\sigma^{B1}-\tilde S_\rQCD(k)d\sigma_B,\quad\ldots\label{subp12}\end{equation}
where the $\bar\beta_n$ are the QCD hard gluon residuals defined above; they
are the non-Abelian analogs of the hard photon residuals defined by YFS.
Here, for illustration, we have recorded the
relationship between the $\bar\beta_n$, $n=0,1$ through ${\cal O}(\alpha_s)$
and the exact one-loop and single bremsstrahlung cross sections,
$d\sigma^{\rm(1-loop)}$, $d\sigma^{B1}$, respectively, where the latter
may be taken from Ref.~\cite{qqOalphas}
We stress two things about the right-hand side of
(\ref{subp10}) :
\begin{itemize}
\item It does not depend on the dummy parameter $K_{max}$ which has been
introduced for cancellation of the infrared divergences in 
$SUM_{IR}(QCD)$ to all orders in $\alpha_s(Q)$ where $Q$ is the hard
scale in the parton scattering process under study here.
\item Its analog can also be derived in our new CEEX~\cite{ceex:2001} 
format.
\end{itemize}

We now
return to the property of (\ref{subp10}) that distinguishes it
from the Abelian result derived by Yennie, Frautschi and Suura 
-- namely, the fact
that, owing to its non-Abelian gauge theory origins, it is in general
expected 
that there are infrared divergences in the $\bar\beta_n$ which were not
removed into the $S_{QCD},\tilde S_{QCD}$ when these infrared functions
were isolated in our derivation of (\ref{subp10}).}\par

More precisely, the left-hand side of (\ref{subp10}) is the fundamental
reduced parton cross section and it should be infrared finite or else
the entire QCD parton model has to be abandoned. 

There is an observation
in the literature~\cite{chris} that 
unless we use the approximation of massless incoming quarks,
the reduced parton cross section
on the left-hand side of (\ref{subp10}) diverges in the infrared
regime at ${\cal O}(\alpha_s^2(Q))$. 
We do not
go into this issue here but either
use the quark masses strictly as collinear limit regulators
so that they are set to zero in the numerators of all
Feynman diagrams in such a way that the limit
$\lim_{m_q^2/E_q^2\rightarrow 0}$, where $E_q$ is the quark
energy, is taken everywhere that
it is finite or, alternatively, we use
n-dimensional methods
to regulate such divergences while setting the quark masses
to zero as that is an excellent approximation for the
light quarks at FNAL and LHC energies -- we take this issue up elsewhere.\par

From the infrared finiteness of the left-hand side of (\ref{subp10})
and the infrared finiteness of $SUM_{IR}(QCD)$, it follows
that the quantity 
\[d\bar{\hat\sigma}_{\rm exp}\equiv 
e^{\rm -SUM_{IR}(QCD)}d\hat\sigma_{\rm exp}\]
must also be infrared finite to all orders in $\alpha_s$.

As we assume the QCD theory makes sense in some neighborhood of the
origin for $\alpha_s$, we conclude that each order in $\alpha_s$
must make an infrared finite contribution to $d\bar{\hat\sigma}_{\rm exp}$. 
At ${\cal O}(\alpha_s^0(Q))$ , the only contribution to 
$d\bar{\hat\sigma}_{\rm exp}$ is the respective Born cross section
given by $\bar\beta^{(0)}_0$ in (\ref{subp10}) and it
is obviously infrared finite, where we use henceforth the notation
$\bar\beta^{(\ell)}_n$ to denote the ${\cal O}(\alpha_s^{\ell}(Q))$
part of $\bar\beta_n$. Thus, we conclude that
the lowest hard gluon residual $\bar\beta^{(0)}_0$ is infrared finite.

Let us now define the left-over 
non-Abelian infrared divergence part
of each contribution $\bar\beta^{(\ell)}_n$ via
\[ \bar\beta^{(\ell)}_n= \tilde{\bar\beta}^{(\ell)}_n + D\bar\beta^{(\ell)}_n\]
where the new function $\tilde{\bar\beta}^{(\ell)}_n$ is now completely
free of any infrared divergences and the function $D\bar\beta^{(\ell)}_n$
contains all left-over infrared divergences in $\bar\beta^{(\ell)}_n$
which are of non-Abelian origin and is normalized to vanish in the
Abelian limit $f_{abc}\rightarrow 0$ where $f_{abc}$ are the group
structure constants.

Further, we define $D\bar\beta^{(\ell)}_n$
by {\it a} minimal subtraction of the respective IR divergences in it
so that it only contains the actual pole and transcendental
constants, $1/\epsilon -C_E$ for $\epsilon=2-d/2$, where $d$ is
the dimension of space-time, in dimensional regularization
or $\ln \lambda^2$ in the gluon mass regularization. Here, $C_E$ is
Euler's constant. 

For definiteness, we 
write this out explicitly as follows:
\[\int dPh\; D\bar\beta^{(\ell)}_n\equiv \sum_{i=1}^{n+\ell} d^{n,\ell}_i\ln^i(\lambda^2)\] 
where the coefficient functions $d^{n,\ell}_i$ are independent of $\lambda$ for
$\lambda\rightarrow 0$ and $dPh$ is the respective n-gluon Lorentz
invariant phase space.

At ${\cal O}(\alpha_s^n(Q))$, the IR finiteness
of the contribution to $d\bar{\hat\sigma}_{\rm exp}$ then requires
the contribution {\small
\begin{eqnarray}
d\bar{\hat\sigma}^{(n)}_{\rm exp} \equiv
\int\sum_{\ell=0}^n\frac{1}{\ell!}\prod_{j=1}^{\ell}\int_{k_j\ge K_{max}}{d^3
k_j\over k_j}\tilde S_{QCD}(k_j)\sum_{i=0}^{n-\ell}\frac{1}{i!}\prod_{j=\ell+1}^{\ell+i}
\nonumber\\
\int{d^3k_j\over k^0_j}
\bar\beta^{(n-\ell-i)}_i(k_{\ell+1},\ldots,k_{\ell+i}){d^3p_2\over p_2^{\,0}}{d^3q_2\over
q_2^{\,0}}\label{subp13}
\end{eqnarray}}
to be finite. 

From this it follows that {\small
\begin{eqnarray}
Dd\bar{\hat\sigma}^{(n)}_{\rm exp} \equiv
\int\sum_{\ell=0}^n\frac{1}{\ell!}\prod_{j=1}^{\ell}\int_{k_j\ge K_{max}}{d^3
k_j\over k_j}\tilde S_{QCD}(k_j)\sum_{i=0}^{n-\ell}\frac{1}{i!}\prod_{j=\ell+1}^{\ell+i}
\nonumber\\
\int{d^3k_j\over k^0_j} D\bar{\beta}^{(n-\ell-i)}_i(k_{\ell+1},\ldots,k_{\ell+i}){d^3p_2\over p_2^{\,0}}{d^3q_2\over
q_2^{\,0}}\label{subp14}
\end{eqnarray}}
is finite. Since the integration region for the final particles is
arbitrary, the independent powers of the IR regulator $\ln(\lambda^2)$ in this
last equation must give vanishing contributions.
This means that
we can drop the $D\bar\beta^{(\ell)}_n$ from our result
(\ref{subp10}) because they do not make a net contribution to the final
parton cross section $\hat\sigma_{\rm exp}$. We thus finally arrive
at the new rigorous result{\small
\begin{equation}
\begin{split}
d\hat\sigma_{\rm exp}&= \sum_n d\hat\sigma^n \\
         &=e^{\rm SUM_{IR}(QCD)}\sum_{n=0}^\infty\int\prod_{j=1}^n{d^3
k_j\over k_j}\int{d^4y\over(2\pi)^4}e^{iy\cdot(p_1+q_1-p_2-q_2-\sum k_j)+
D_\rQCD}\\
&*\tilde{\bar\beta}_n(k_1,\ldots,k_n){d^3p_2\over p_2^{\,0}}{d^3q_2\over
q_2^{\,0}}
\end{split}
\label{subp15}
\end{equation}
where now the hard gluon residuals 
$\tilde{\bar\beta}_n(k_1,\ldots,k_n)$
defined by 
\begin{equation}
\tilde{\bar\beta}_n(k_1,\ldots,k_n= \sum_{\ell=0}^\infty 
\tilde{\bar\beta}^{(\ell)}_n(k_1,\ldots,k_n)
\label{newbeta}
\end{equation}
are free of all infrared divergences to all 
orders in $\alpha_s(Q)$.
This is a basic result of this Appendix.\par
 
We note here that, contrary to what was claimed in the Appendix of the
first paper in Refs.~\cite{delaney} and consistent with what is
explained in the third reference in ~\cite{delaney}, the arguments in 
the first paper in Refs.~\cite{delaney}
are not sufficient to derive the respective analog of eq.(\ref{subp15});
for, they did not really expose the compensation between
the left over genuine non-Abelian IR virtual and real singularities
between $\int dPh\bar\beta_n$ and $\int dPh\bar\beta_{n+1}$ respectively
that really distinguishes
QCD from QED, where no such compensation occurs in the $\bar\beta_n$
residuals for QED.\par

We point-out that the general non-Abelian exponentiation of the 
eikonal cross sections in QCD has been proven formally in 
Ref.~\cite{gatherall}. The contact between Ref.~\cite{gatherall}
and our result (\ref{subp15}) is that, in the language of
Ref.~\cite{gatherall}, our exponential factor corresponds to the 
N=1 term in the exponent of eq.(10) of the latter reference.
One also sees immediately the fundamental difference between what we
derive in (\ref{subp15}) and the eikonal formula in Ref.~\cite{gatherall}:
our result (\ref{subp15}) is an {\it exact} re-arrangement 
of the complete cross section
whereas the result in eq.(10) of Ref.~\cite{gatherall} is an
{\it approximation} to the complete cross section in which all terms that
could not be eikonalized and exponentiated have been dropped.
\par


\end{document}